\documentclass[journal]{IEEEtran}
\ifCLASSINFOpdf
\else
\fi
\usepackage[T1]{fontenc} % optional
\usepackage{amsmath}
\usepackage{bm} % optional
\usepackage{cite}
\ifCLASSOPTIONcompsoc
\usepackage[caption=false,font=normalsize,labelfont=sf,textfont=sf]{subfig}
\else
\usepackage[caption=false,font=footnotesize]{subfig}
\fi

\usepackage{amssymb}
\usepackage{graphicx}% Include figure files
\usepackage{float} %Force place figure
\usepackage{dcolumn}% Align table columns on decimal point
\usepackage{verbatim}   % useful for program listings
\usepackage{color}      % use if color is used in text
\usepackage{siunitx} %use for type number and SI units
\usepackage[capitalise]{cleveref} %use cref{label} to generate automated ref. of single or multiple eq/fig.
\usepackage{verbatim} %multiline comment out

\begin{document}
\title{Quantum Size Effects in the Terahertz Nonlinear Response of Metallic
Armchair Graphene Nanoribbons}

\author{Yichao~Wang
        and~David R.~Andersen
\thanks{Yichao Wang is with the Department
of Electrical and Computer Engineering, University of Iowa, Iowa City,
IA, 52242 USA. (e-mail: yichao-wang@uiowa.edu).}% <-this % stops a space
\thanks{David R. Andersen is with the Department
of Electrical and Computer Engineering and the Department of Physics and Astronomy, University of Iowa, Iowa City,
IA, 52242 USA. (e-mail: k0rx@uiowa.edu).}% <-this % stops a space
}

\maketitle

\begin{abstract}
We use time dependent perturbation theory to study quantum size effects on the
terahertz nonlinear response of metallic graphene armchair nanoribbons of
finite length under an applied electric field. Our work shows that
quantization due to the finite length of the nanoribbon, the applied field
profile, and the
broadening of the graphene spectrum all play a significant role in the resulting
nonlinear conductances. In certain cases, these effects
can significantly enhance the nonlinearity over that for
infinitely-long metallic armchair graphene nanoribbon.
\end{abstract}

%\begin{IEEEkeywords}
%IEEE, IEEEtran, journal, \LaTeX, paper, template.
%\end{IEEEkeywords}

% \fi
\IEEEpeerreviewmaketitle

\section{Introduction}
\IEEEPARstart{G}{raphene} has many unique electronic, mechanical, thermal and optoelectronic
properties\cite{grreview}. A tunable Fermi level and linear dispersion relation
near the Dirac point are some of the features that make graphene attractive for
the study of nonlinear effects in the terahertz (THz)
regime\cite{nlr07,neto09,coherent10,sarma11,4wavegr12,review2014}. Various
theoretical predictions of the generation of higher-order harmonics in graphene
structures had been performed
\cite{mikhailov2008nonlinear,mikhailov2012theory,wright09,blg10,subgap11,gullans2013single}.
Recent experimental reports on the measurement of the THz nonlinear response in
single- and multi-layer graphene\cite{maeng2012gate,kumar2013third,noepgr15}
further demonstrate that graphene structures possess
a strong nonlinear THz response. These theoretical and experimental studies
demonstrate that, when compared to conventional parabolic semiconductor
structures, unique graphene properties, such as linear energy dispersion, high
electron Fermi velocity and tunable Fermi level, lead to a stronger nonlinear
optical response in many 2D graphene structures \cite{nlr07,neto09,coherent10,sarma11,4wavegr12,review2014,mikhailov2008nonlinear,mikhailov2012theory,wright09,blg10,subgap11,gullans2013single,maeng2012gate,kumar2013third,noepgr15}.\par

Unlike the extensive research on the nonlinear response of 2D graphene
structures, prior to our work, only the linear THz response
\cite{duan2010infrared,liao08optical} and selection rules
\cite{gnrselectionrule,gnrselectionrule2} of graphene nanoribbons (GNR) for a
linearly polarized electric field have been investigated. %Recently, theoretical and experimental studies also demonstrated the nonlinear plasmon existed in GNR structures \cite{nasari16nonlinear,cox15plasmon}. 
Thin GNRs (sub-\SI{20}{nm}) with smooth edges can be treated as quantum wires,
not dominated by defects \cite{GNRisQW}. The reduced dimensionality of 2D
graphene to a quasi 1D quantum wire for narrow GNR opens the study of new physics
(including quantization of energy, momentum \textit{etc.}). GNRs have two types
of edges: armchair graphene nanoribbons (acGNR) and zigzag graphene nanoribbons
(zzGNR). These two types of GNR shows distinct electronic characteristics due to
the geometry and boundary conditions\cite{fertig1,fertig2,and1,and2}. In
general,  redistribution of the Dirac fermions  induced by the applied electric
field in momentum and energy space leads to large THz nonlinearities in GNR. The resulting nonequilibrium distribution predicts the conductivity components oscillating in time and space, and spatially homogeneous steady state components. As a result, nonlinear response in GNRs are sensitive to the applied field strength and polarization\cite{review2014}. 

A widely used model, the perturbation of the Fourier expansion of the wave
function first adopted by Wright \textit{et.al.} \cite{wright09} in the study of
the THz nonlinear response of various 2D graphene systems
\cite{wright09,blg10,subgap11,ang2015nonlinear} has two important assumptions:
\textit{i)} the absence of coupling between the induced nonlinear response and
the applied electric field spatial profile; and \textit{ii)} charge carriers
propagate with ideal ballistic transport in graphene, with the absence of
broadening due to various scattering processes \cite{arxiv1}. Velicky,
Ma{\v{s}}ek and Kramer have developed a model of the AC
ballistic/quasi-ballistic conductance in 1D quantum wires with an arbitrary
spatial profile of the applied electric
field\cite{mavsek89,velicky89ac,kramer90coherent}. Wr{\'o}bel \textit{et.al.}
measured and analyzed the role of reduced dimensionality in the quantized
conductance of an GaAlAs/GaAs quantum wire
\cite{wrobel92quantized,wrobel94influence}. As thin GNRs with finite length in
our study possess a low dimensional mesoscopic structure, it is natural to use
these ideas to extend this analytical approach to the nonlinear response of thin
GNRs with an applied electric field.\par

Carrier relaxation in graphene near the Dirac point is caused primarily by
scattering of hot carriers
\cite{tau25ps,strait11cool,johannsen13direct,gierz13snapshots}. Two typical
carrier relaxation times have been reported, \SI{25}{ps} for
$E_F\ll\SI{200}{meV}$, where \SI{200}{meV} corresponds to the optical phonon
energy \cite{tau25ps} and a few \si{ps} for states involving optical phonon scattering \cite{george08,tau25ps,strait11cool,johannsen13direct,gierz13snapshots,jnawali13}.  
Current limitations to the utilization of thin GNRs for nonlinear device
applications result from scattering due to edge defects and hot carriers
\cite{yang10impact,liao11thermally,strait11cool,finite12,johannsen13direct,gierz13snapshots}.
Furthermore, edge disorder can affect the interband transition process due to
the extra energy required to satisfy the conservation laws in the interband
transition process \cite{george08,romanets10}. Our work focuses on the THz emission
due to direct interband transition of graphene carriers with energy
$\ll\SI{200}{meV}$ in thin metallic acGNR. Scattering due to hot carriers and optical phonons 
is reduced for the THz direct interband transition
\cite{tse09,romanets10,johannsen13direct}. Further, scattering due to acoustic phonons is
prohibited for these interband transition \cite{tse09}. Therefore, carrier
relaxation in finite GNR structures mainly depends on edge disorder and defects.\par

Theoretical studies show that non-perfect edges destroy the quantization of the
conductance for GNRs \cite{edge15review}. However, the rapid development of
techniques for the synthesis of thin GNRs 
\cite{GNRisQW,kimouche2015ultra,jacobberger2015direct}, show that thin GNR
may have ultra smooth edges, higher mobility and longer carrier mean free path
than expected theoretically. The recent reported synthesis of ultra thin acGNR
(sub-\SI{10}{nm}) show that the electronic structure of ultrathin acGNR is not
strongly affected by defects (kinks)
\cite{kimouche2015ultra,jacobberger2015direct}.  It is possible for thin GNR
mesoscopic structures grown in the laboratory to show ballistic and
quasi-ballistic transport. Scattering along the channel direction is greatly
reduced in the ballistic and quasi-ballistic regime\cite{GNR15review}. Such progess in the state
of art of the growth of ultra thin GNR highlights the potential for quasi 1D GNR
mesoscopic structures to be used in modern ultra-high-speed electronic and quantum devices \cite{GNR15review}. Thus the study the nonlinear electrodynamics for thin metallic acGNR with an applied electric field with finite length in the mesoscopic regime is of particular significance today.\par

In this paper, we present important new results showing
that the quantum size effects of nanonribbon length, spectral
broadening, and excitation field coupling significantly modify
the THz nonlinear response of thin metallic acGNR. {\it These
novel effects play an essential role in the behavior of THz
nonlinearities in acGNR, and have not previously been investigated.}
In particular, we find that quantization of the broadened
Dirac particle spectrum results in a transition from a discrete
quantum dot-like spectrum for small nanoribbon lengths to a
continuous spectrum as the length of the nanoribbon increases.
We evaluate the boundary between these two qualitatively
distinct behaviors in terms of the coupling between adjacent
energy states due to the broadening. Further, we find that the
exact spatial profile of the THz excitation field plays a
significant role in the nonlinear response. The spatial Fourier
spectrum of the excitation field serves to enhance the nonlinearity
at photon energies near states where the spectrum exhibits
maxima, and reduces the response near spectral minima.
By apodizing \cite{bornwolf} the excitation field profile it becomes
possible to optimize the THz nonlinearities at a particular
desired pump frequency.\par

The paper is organized as follows: In Sec. \ref{Model}, we
model the THz nonlinear response for thin metallic acGNR of finite length.
We analyze the nonlinear response of these acGNR in the presence of intrinsic
broadening and the coupling of the applied electric field profile to study
the impact of quantum size effects on the THz nonlinearities.
In Sec. \ref{Results}, we apply our model to calculate the
nonlinear THz conductance of thin metallic acGNR. We analyze the
dependence of the third-order nonlinear terms on the ribbon length, temperature,
and length of illumination. Following the introduction of broadening, we propose
an effective critical length, characterizing the quantization of energy due to
finite length impacting the continuum of states. We then
show that in metallic acGNR with length smaller than the effective critical
length, the THz third-harmonic conductance is greatly enhanced to nearly the
order of the THz third-order Kerr conductance in the THz regime. This result
shows that the tunability of thin metallic acGNR in the terahertz regime is
increased. Finally, we present our conclusions in Section \ref{Conclusions}.

\section{Model}\label{Model}
Following the  low energy model for GNR\cite{fertig1,fertig2}, the time-dependent, unperturbed
$\mathbf{k} \cdot \mathbf{p}$ Hamiltonian for a single Dirac fermion near the Dirac points may be written in terms of Pauli matrices as $H_{0,K} = \hbar v_F \bm{\sigma} \cdot \mathbf{k}$ for the $\mathbf{K}$ valley
and $H_{0,K'} = \hbar v_F \bm{\sigma} \cdot \mathbf{k'}$ for the $\mathbf{K'}$
valley with $\mathbf{k}\, (\mathbf{k'})$ the perturbation from the center of
the $\mathbf{K} \, (\mathbf{K'})$ valley. The time-independent (unperturbed)
Hamiltonian for GNR may be written:
%\begin{equation}
%\begin{aligned}
%H_0&=\begin{pmatrix} H_{0,K}&0\\0&H_{0,K'}\end{pmatrix}\\
%&=\hbar  v_F
%\begin{pmatrix}
% 0 & k_x-i k_y & 0 & 0 \\
% k_x+i k_y & 0 & 0 & 0 \\
% 0 & 0 & 0 & -k_x-i k_y \\
% 0 & 0 &-k_x+i k_y & 0  \end{pmatrix}
%\end{aligned}
%\end{equation}
\begin{equation}
\begin{aligned}
H_0&=\hbar  v_F
\begin{pmatrix}
 0 & k_x-i k_y & 0 & 0 \\
 k_x+i k_y & 0 & 0 & 0 \\
 0 & 0 & 0 & -k_x-i k_y \\
 0 & 0 &-k_x+i k_y & 0  \end{pmatrix}
\end{aligned}
\end{equation}
with wave functions in the case of acGNR:
\begin{equation}\label{eq:psi0}\psi_{n,s}=\frac{e^{i k_y y}}{2\sqrt{L_x L_y}}
\begin{pmatrix} e^{\textnormal{-} i \theta_{k_n,k_y}} e^{i k_n x}\\s e^{i k_n
x}\\-e^{\textnormal{-} i \theta_{k_n,k_y}} e^{\textnormal{-}i k_n x}\\s
e^{\textnormal{-}i k_n x}\end{pmatrix}
\end{equation}\\
where $L_x=N a_0/2$ is the width of the acGNR in the $\hat{x}$ (zigzag) direction,  $L_y$
is the length of the acGNR in
the $\hat{y}$ (armchair) direction, and $\theta_{k_n,k_y}=\tan^{-1} (k_n/k_y)$ is the
direction of the isospin state. This Hamiltonian does not include intervalley
scattering processes due to its block-diagonal character.\par

The width of acGNR determines the
metallic or semiconductor character of the acGNR\cite{fertig1,fertig2}. In
general, acGNR of $N=3M-1$ atoms wide along the zigzag edge, with \textit{M}
odd, are metallic, whereas all other cases are semiconducting.
The energy dispersion relation arising from this model is doubly-degenerate, with one
branch coming from each of the $\mathbf{K}$ and $\mathbf{K'}$ valleys.\par

\subsection{AC conductance}
Due to the quasi-1D structure of the thin acGNR \cite{GNRisQW} and the resultant quantization in $k$
space, we need to consider the coupling of the applied electric field with the
quantized $k$-states. The AC
conductance $\tilde{g}(\omega)$ is defined in terms of the absorbed power
$P(\omega)$ for
an acGNR locally excited with electric field $\mathbf{E}(\mathbf{r},\omega)$:
\begin{equation}
\tilde{g}(\omega)\equiv \frac{P(\omega)}{\phi^2(\omega)/2}, \hspace{0.4cm}
\phi(\omega)=\int \hspace{0.05cm} \mathbf{E}(\mathbf{r},\omega)
\cdot d\mathbf{r}
\end{equation}
where $\phi(\omega)$ is the change of the electric potential in the irradiated
region. The absorbed power may be expressed by the conductivity and the acting
field as:
\begin{equation}
\tilde{g}(\omega)=\frac{1}{2}\int \frac{dk}{2\pi} \hspace{0.05cm} \sigma(k,\omega) \left | E(k,\omega) \right |^2\end{equation}
The \textit{i}th order AC conductance for infinitely long acGNR ${\tilde{g}_{y
\nu}}^{(i)}(\omega)$ is written\cite{mavsek89,velicky89ac,kramer90coherent}:
\begin{equation} \label{eq:gw}
{\tilde{g}_{y\nu}}^{(i)}(\omega,L)=\frac{g_s g_v L_y}{2\pi} \int_{-\infty}^{\textnormal{+}\infty} dk_y \hspace{0.05cm} \sigma_{y\nu}^{(i)}(m,\omega) \left (  \frac{\sin(k_y L/2)}{k_y L/2}\right )^2
\end{equation}
with $L$ the length of illumination. For simplicity, we assume a constant
field strength over the length of illumination $L$.\par

Defining  $\omega_y = k_y v_F$,  the corresponding angular frequency of $k_y$
in GNR with a group velocity of $v_F$ in the relaxation-free approximation, neglecting all scattering effects \cite{arxiv1}, we rewrite \eqref{eq:gw} for the third-order AC conductance as:
\begin{subequations}\label{eq:gwy}
\begin{align} 
{\tilde{g}_{y \nu}}^{(3)}(\omega,L)&=\sum_{l=1}^{2}2\int_{0}^{\infty} d\omega_y
\, {\tilde{f}_{y\nu}}^{(\omega)}(\omega_y,\frac{\omega l}{2}) \delta(\omega_y-\frac{\omega l}{2}) \label{eq:gwy1}\\
{\tilde{g}_{y \nu}}^{(3)}(3\omega,L)&=\sum_{l=1}^{3}2\int_{0}^{\infty} d\omega_y
\, {\tilde{f}_{y\nu}}^{(3 \omega)}(\omega_y,\frac{\omega l}{2}) \delta(\omega_y-\frac{\omega l}{2})  \label{eq:gwy3}
\end{align}
\end{subequations}
where \begin{equation}
{\tilde{f}_{y
\nu}}^{(\omega_0)}(\omega_{y},\frac{\omega l}{2})=f_{y\nu}^{(\omega)}(\omega_y,\frac{\omega l}{2}) N(\omega_y) S(\omega_y,L)
\end{equation} 
with the thermal factor:
\begin{equation}
N (\omega_y)=\frac{\sinh \left( \frac{\hbar |\omega_y|}{k_B T} \right)}{\cosh \left( \frac{E_F}{k_B T} \right)+\cosh \left( \frac{\hbar |\omega_y|}{k_B T} \right)}
\end{equation}
and the illumination factor:
\begin{equation}
S( \omega_y,  L) = \mathrm{sinc}^2\left( \frac{\omega_y L}{2 v_F} \right)
\end{equation}
and where $f_{y\nu}^{(\omega_0)}(\omega_y,\frac{\omega l}{2})$ is the
coefficient associated with the
corresponding $\delta(\omega_y-\frac{\omega l}{2})$ term in the expansion of the
expression for the local third-order conductivity (see \cite[eq. (41-42)]{arxiv1}).\par

With an applied electric field linearly polarized along the $\hat {y}$ direction
of an infinitely long acGNR, for the metallic band where $k_{x,n}=0$, the AC isotropic nonlinear conductance becomes\cite{arxiv1}: 
\begin{subequations}
\label{eq:gACy}
\begin{align}
{\tilde{g}_{yy}}^{(1)}(\omega,L) & = - g_0\eta_{x}  N \left(\frac{\omega}{2}
\right)  S\left( \frac{\omega}{2},L \right)\\
{\tilde{g}_{yy}}^{(3)}(\omega,L) & = - g_0 \eta \,
\eta_{x}\sum_{l=1}^{2}\left(\frac{1}{2}\right)^{-z(l)} N \left(\frac{\omega
l}{2} \right) S\left( \frac{\omega l}{2},L \right)\\
{\tilde{g}_{yy}}^{(3)}(3\omega,L) & =-g_0 \eta\, \eta_{x}\sum_{l=1}^{3}
\left(-\frac{1}{2}\right)^{z(l)} N \left(\frac{\omega l}{2} \right) S\left(
\frac{\omega l}{2},L \right)
\end{align}
\end{subequations}
similarly, the AC anisotropic nonlinear conductance is:
\begin{subequations}
\label{eq:gACx}
\begin{align}
{\tilde{g}_{yx}}^{(1)}(\omega,L) & =g_0\eta_{x}  N \left(\frac{\omega}{2}
\right) S\left( \frac{\omega}{2},L \right)\\
{\tilde{g}_{yx}}^{(3)}(\omega,L) & =g_0 \eta\, \eta_{x} N(\omega)  S\left(
\omega, L \right )\\
{\tilde{g}_{yx}}^{(3)}(3\omega,L) & =g_0 \eta\, \eta_{x}\sum_{l=1}^{3}
\left(-\frac{1}{2}\right)^{z(l)} N \left(\frac{\omega l}{2} \right) S\left(
\frac{\omega l}{2},L \right)
\end{align}
\end{subequations}
where the quantum conductance $g_0=e^2/\left(4\hbar^2\right)$, Fermi level
$E_F=h f_{\mu}$, harmonic constant $z(l)=\left[1-(-1)^l\right]/2$,  gain due to
the width $\eta_{x}=\left(g_s g_v v_F\right)/\left(\omega L_x\right)$ and the
coupling strength
$\eta=\left(e^2E_y^2v_F^2\right)/\left(\hbar^2\omega^4\right)$. The illumination
factor $S(\omega,L)$ in \eqref{eq:gACy} and \eqref{eq:gACx} arises from the
finite illumination length and is the square modulus of the Fourier transform of
the applied field profile. As a result of the inversion symmetry inherent in acGNR, the \textit{2}nd-order current makes no contribution to the total current.\par

The total third-order nonlinear conductance for metallic acGNR then can be
expressed as:
\begin{equation}\label{eq:g3yy}
{\tilde g_{tot,y \nu}}^{(3)}(\omega, L) ={\tilde g_{y \nu}}^{(3)}(\omega,L)
e^{-i \omega
t}+{\tilde g_{y \nu}}^{(3)}(3\omega,L)e^{-i 3 \omega t} + c.c. 
\end{equation}
This result shows that for infinitely long metallic acGNR, the third-order
nonlinear conductance is a superposition of two frequency terms: \textit{i)}
${\tilde g_{y\nu}}^{(3)}(\omega,L)$, the Kerr conductance term
corresponding to the absorption of two photons and the simultaneous emission of
one photon; and \textit{ii)} $g_{y\nu}^{(3)}(3 \omega,L)$, the third-harmonic conductance
term corresponding to the simultaneous absorption of three
photons. The complex conjugate parts in \eqref{eq:g3yy} are for
the emission process.\par

We observe that by taking the limit $L \rightarrow 0$, the ideal conductance and AC
conductance in our definition are equivalent: $g_{\mu
\nu}^{(i)}(\omega)={\tilde{g}_{\mu \nu}}^{(i)}(\omega,0)$. For $L=0$, there is
no coupling between the induced nonlinear response and the applied field
spatial profile. Due to the current operator $q v_F \sigma_{x,y}
\delta(\mathbf{r}-\mathbf{r_{op}})$ used in our previous
work\cite{arxiv1}, we assume graphene carriers at $\mathbf{r_{op}}$
interact only with the incoming photon field at $\mathbf{r_{op}}$. The
conductance $g_{\mu \nu}^{(i)}(\omega)$ is a special case of the AC conductance
${\tilde{g}_{\mu \nu}}^{(i)}(\omega,L)$, and the AC conductance ${\tilde{g}_{\mu
\nu}}^{(i)}(\omega,L)$ reduces to the ideal conductance $g_{\mu
\nu}^{(i)}(\omega)$ at $L = 0$.\par

\subsection{Broadening}
We employ a Gaussian broadening model to study the impact on the nonlinear
conductance due to spectral broadening of acGNR in the THz regime. This Gaussian broadening method has been
widely used in the study of many graphene and GNR structures \cite{rotenberg2008origin,xia2011origins,chen2015molecular}. The
Gaussian kernel,
\begin{equation}\label{eq:gaussian}
Z_g(\omega_y,\omega)=\frac{1}{\sqrt{\pi}\Gamma_{\omega}}\exp \left[-\frac{(\omega_y-\omega)^2}{\Gamma_{\omega}^2}\right]
\end{equation}
with $\Gamma_{\omega}=2\pi f_{\Gamma}=2\pi\left(2\tau \sqrt{\ln 2}  \right)^{-1}$ and $\tau$ the relaxation time, replaces the Dirac delta function in the integrand of \eqref{eq:gwy}. \par

In this work, we neglect edge defects in the acGNR. We further assume the
broadening parameter remains a constant in the THz regime, and is invariant of
the temperature and applied field strength $E_y$. We obtain the value
$\tau=\SI{25}{ps}$ from \cite[Table I]{tau25ps}
and therefore, the broadening parameter used in \eqref{eq:gaussian} becomes $f_{\Gamma}=\SI{0.024}{THz}$.
This choice of the broadening parameter is appropriate because the direct
interband transition in the THz regime is well below the \SI{200}{meV} optical
phonon band.  We note however, that our model can be extended to situations with
larger carrier scattering. As $\tau$ is reduced, the quantization of the
conductance tends to be dominated by other scattering processes. As a result,
the mean free path of the carriers becomes shorter, and the interaction between
adjacent states defined by the quantization condition becomes stronger.

\subsection{Quantization due to finite length}
In all real nanoribbons, the length $L_y$ of the nanoribbon will be finite. This
results in a discrete set of electronic states along the $k_y$ direction, as
opposed to the continuum of states that results for $L_y \to \infty$. In this
case, the resulting third-order conductances are obtained by summing over the
discrete set of states rather than integrating over the continuum.

In metallic acGNR, when $k_{x,n}=0$, the energy dispersion relation may be
written $\epsilon=s|m|\hbar \omega_{y0}$, with $\omega_{y0}=2\pi v_F/L_y$. For a
finite nanoribbon length of $L_y$ and broadening $\Gamma_{\omega}$, with $\omega_y=m \omega_{y0}$, the third-order AC conductance becomes:
\begin{subequations}\label{eq:gwyG}
\begin{align} 
{\tilde{g}_{y
\nu}}^{(3)}(\omega,L,L_y,\Gamma_{\omega})&=\sum_{m=0}^{\infty}\sum_{l=1}^{2}2
{\tilde{f}_{y\nu}}^{(\omega)}(\omega_{y},\frac{\omega l}{2}) Z_g(\omega_{y},\frac{\omega l}{2})\omega_{y0} \label{eq:gwyG1}\\
{\tilde{g}_{y
\nu}}^{(3)}(3\omega,L,L_y,\Gamma_{\omega})&=\sum_{m=0}^{\infty}\sum_{l=1}^{3}2
{\tilde{f}_{y\nu}}^{(3 \omega)}(\omega_{y},\frac{\omega l}{2}) Z_g(\omega_{y},\frac{\omega l}{2})\omega_{y0}  \label{eq:gwyG3}
\end{align}
\end{subequations}
%\IEEEpubidadjcol

\section{Results and Discussion}\label{Results}
We consider
thin acGNR with finite length $L_y$, for which there exists an energy
quantization with quantum number $m$, for states of the linear bands near the
Dirac points in thin metallic acGNR. To simplify the discussion, we present
results for acGNR20, the armchair graphene nanoribbon $N=20$ atoms wide, which
can be treated as a quasi 1D quantum wire \cite{GNRisQW}, and the applied
field strength is $E_y=\SI{10}{kV/m}$ throughout.
In what follows, we summarize
the characteristics of the AC nonlinear conductance for all combinations of
length of illumination and Fermi level, given in \eqref{eq:gwyG}.
\begin{figure*}[!t]
\centering
\subfloat{\label{fig:1a}\includegraphics[width=2.5in]{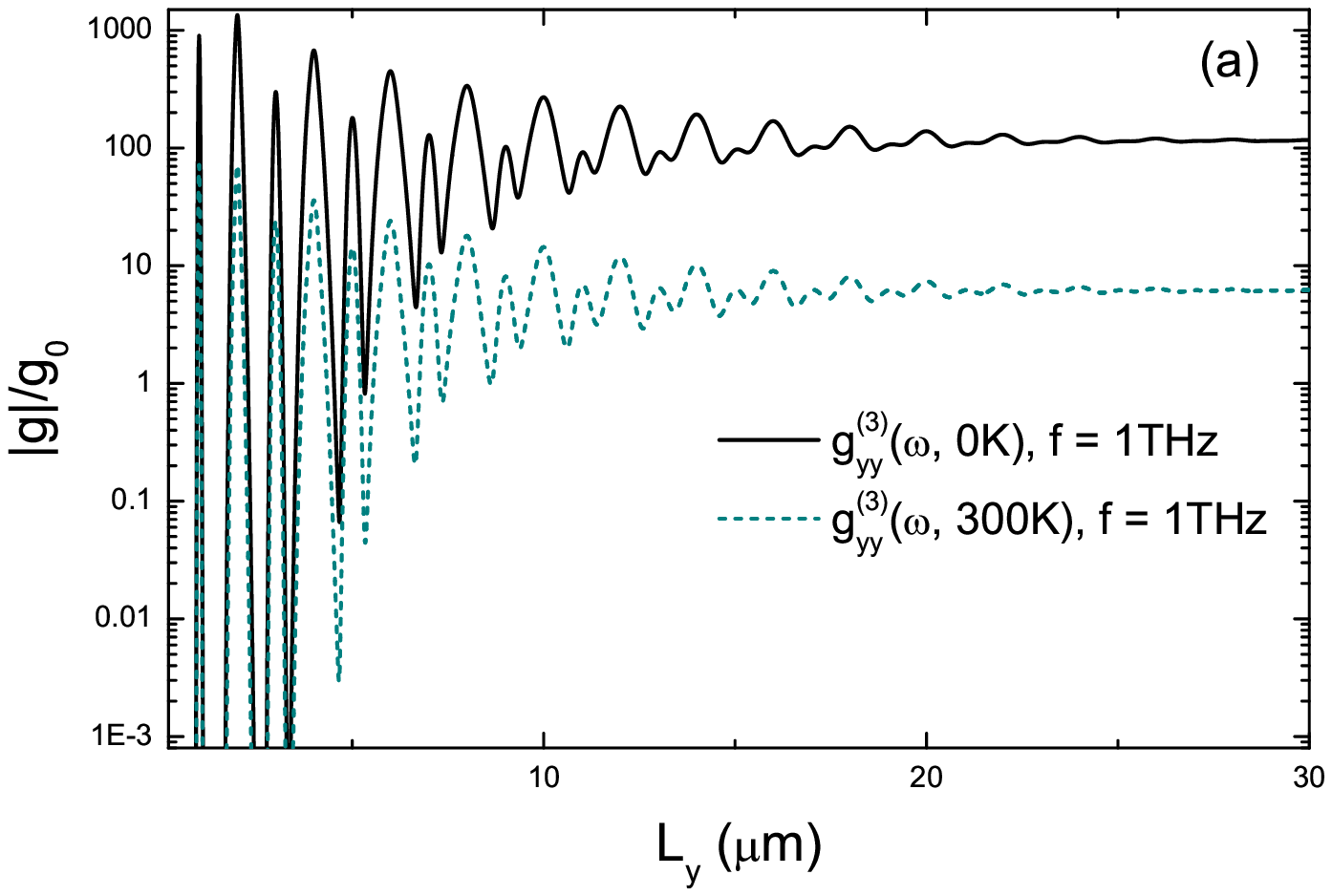}}
\hfil
\subfloat{\label{fig:1b}\includegraphics[width=2.5in]{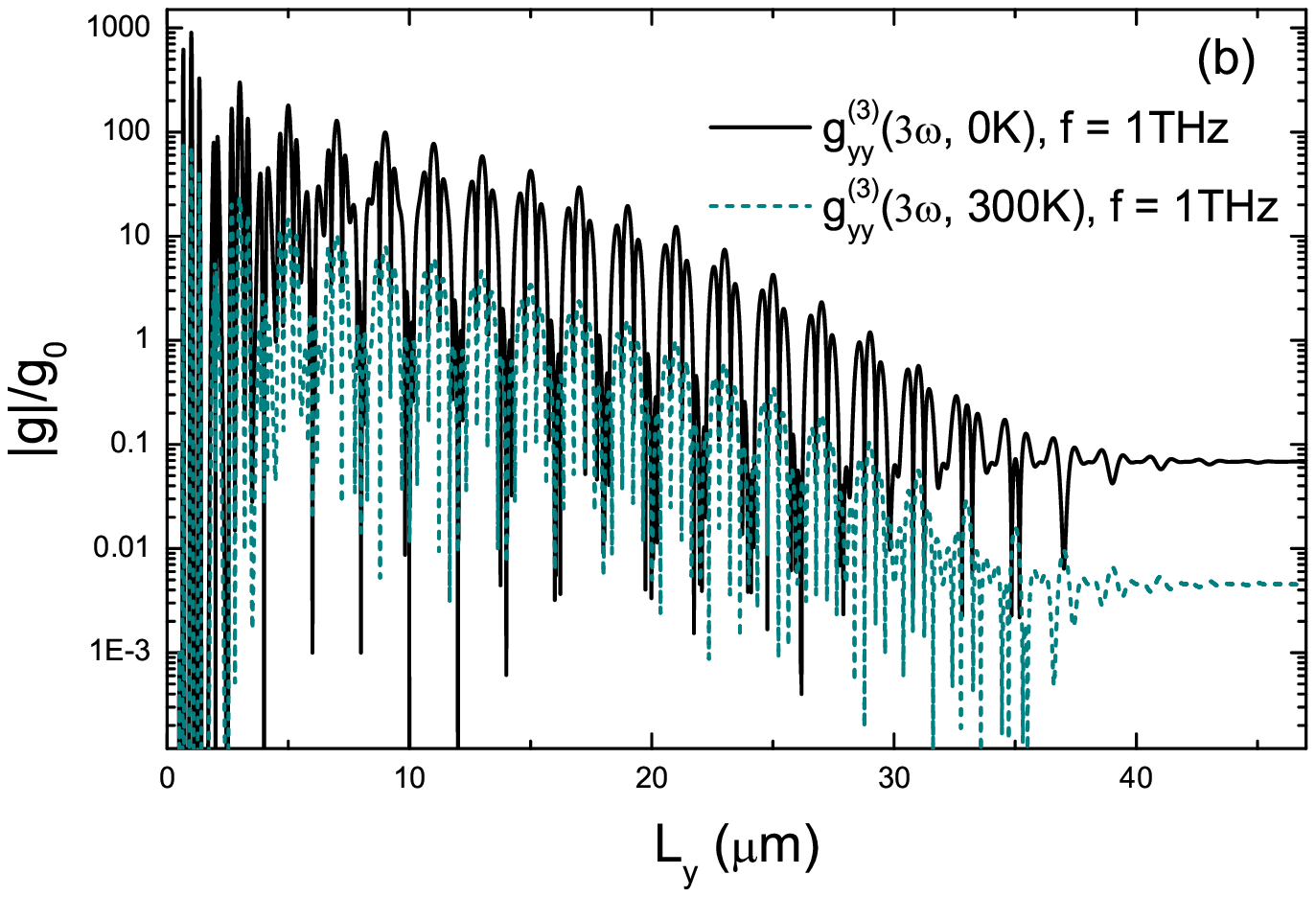}}
\hfil
\subfloat{\label{fig:1c}\includegraphics[width=2.5in]{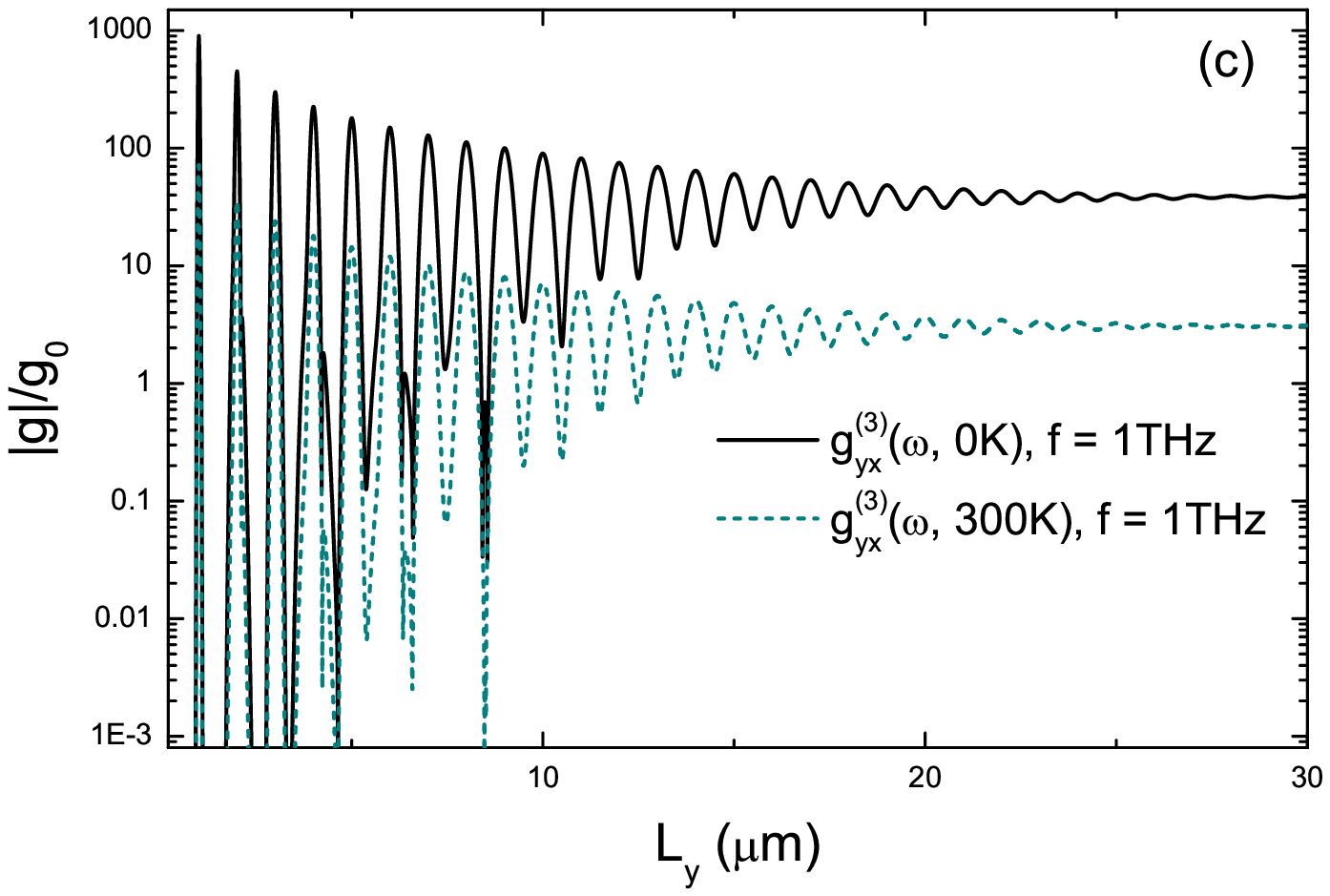}}
\hfil
\subfloat{\label{fig:1d}\includegraphics[width=2.5in]{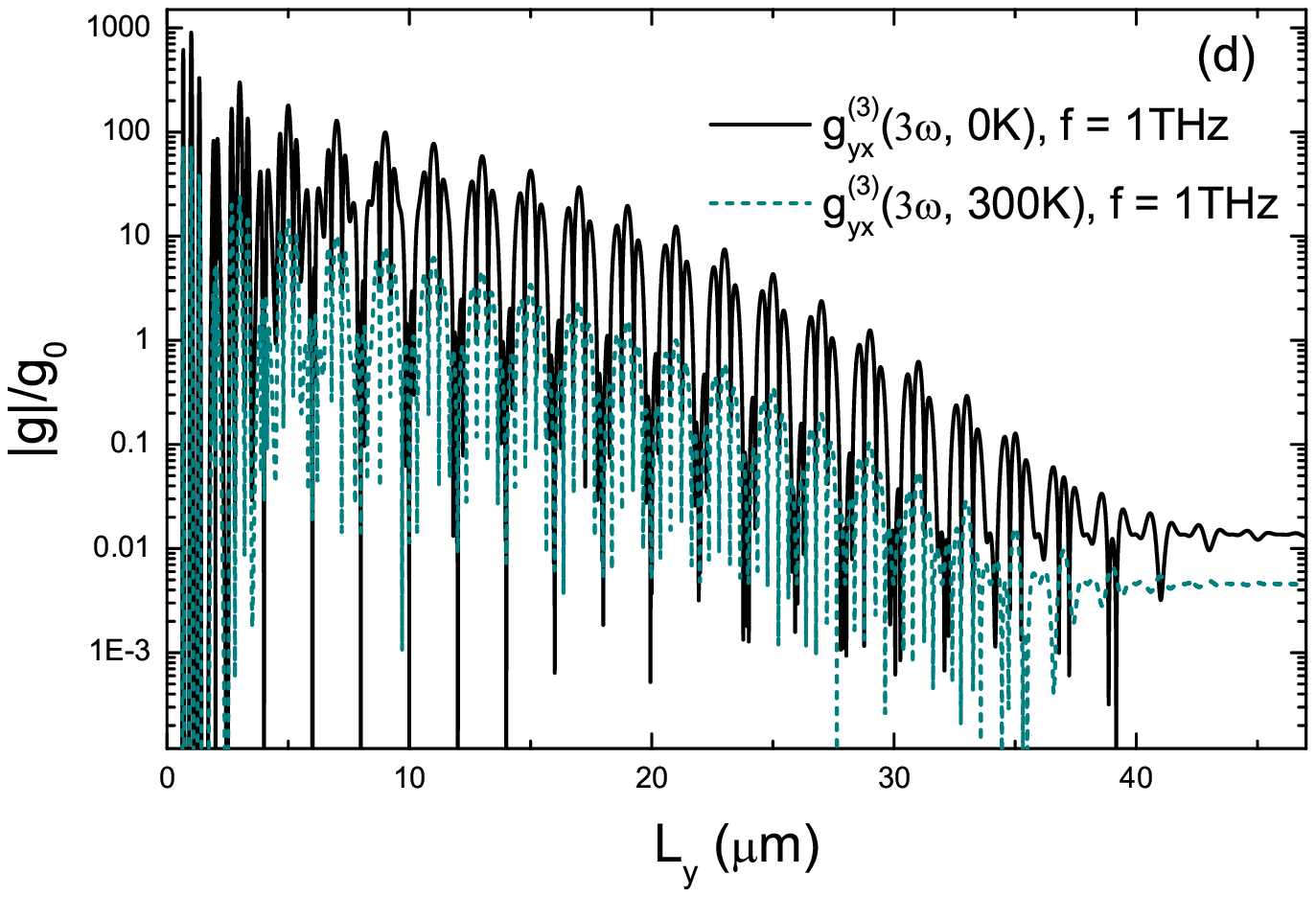}}
\caption{Magnitude of the third-order nonlinear conductances in acGNR20 at 
$T = \SI{0}{K}$ and $\SI{300}{K}$ as a
function of nanoribbon length $L_y$. a) isotropic Kerr conductance,
b) isotropic third-harmonic conductance,
c) anisotropic Kerr conductance,
and d) anisotropic third-harmonic
conductance. For all plots,  $f = \SI{1}{THz}$, $E_F=0$, $L=0$, and
$f_{\Gamma}=\SI{0.024}{THz}$.}
\label{fig:1}
\end{figure*}
We plot the isotropic and anisotropic AC conductances as a function of $L_y$
with $T = \SI{0}{K}$ and \SI{300}{K}, and $L \to 0$ for intrinsic acGNR in Fig. \ref{fig:1}. The
frequency of the applied field is $f =\SI{1}{THz}$. Due to the energy quantization
resulting from finite $L_y$, interband transitions can only be excited for
states coupled by the excitation frequency $2 \pi f = \omega$, namely those
where $\omega/2 = M
\omega_{y0}/2$ (single-photon resonance), $\omega = M \omega_{y0}$ (two-photon
resonance), and $3 \omega/2 = 3 M \omega_{y0} /2$ (three-photon resonance for the
third-harmonic nonlinearity) where $M$ is
an integer and where $\omega_{y0} = 2 \pi v_F
/ L_y$ is the separation between the discrete states. If $M$ is odd, then $M/2$ and
$3M/2$ are not integers and states at these energies do not exist.
This implies that contributions from the
$\omega/2$ and $3\omega/2$ components of the conductance are nearly zero
(exactly zero in the absence of broadening). If $M$ is even,
$M/2$ and $3M/2$ are both integers, thus contributions from the $\omega/2$ and
$3\omega/2$ components of the conductance are non-zero.
In summary, for $M=f
L_y/v_F$ even, \eqref{eq:gwyG} is equivalent to
(\ref{eq:gACy}, \ref{eq:gACx}).
For $M$ odd, the $ N(\omega/2)$ and
$N(3\omega/2)$ terms in \eqref{eq:gwyG} are negligible.
In Fig. \ref{fig:1}, we can see clearly how the ribbon length $L_y$ affects the
conductance. For small $L_y$, the broadening is smaller than the separation
between states and we observe quantization of the conductance.
In essense, the acGNR behaves as a rectangular quantum dot.
When $L_y$ becomes longer, the states move closer together and there is overlap due
to broadening for adjacent energy states. As a result, the overall conductance
approaches a constant value.  We arrive at an effective critical length
$L_{yc}^{(\omega)}=2 v_F /3f_{\Gamma}$ bounding the quantum and continuum
regions for the Kerr conductance. Similarly, the
effective critical length for the third-harmonic conductance
$L_{yc}^{(3\omega)}=v_F /f_{\Gamma}$. Both critical lengths are independent of
$f$ in the THz regime. For $f_{\Gamma}=\SI{0.024}{THz}$,
$L_{yc}^{(\omega)}\sim\SI{28}{\micro m}$ and
$L_{yc}^{(3\omega)}\sim\SI{42}{\micro m}$. When $L_y$ is greater than the
effective critical length, the conductance converges to the conductance of an
infinitely-long acGNR, and we enter the quasi-continuum regime due to the fact
that the broadening now overlaps adjacent states. Such asymptotic behaviour
can be observed in Fig. \ref{fig:1} when $L_y$ is greater than the critical length.\par
\begin{figure*}[!t]
\centering
\subfloat{\label{fig:2a}\includegraphics[width=2.5in]{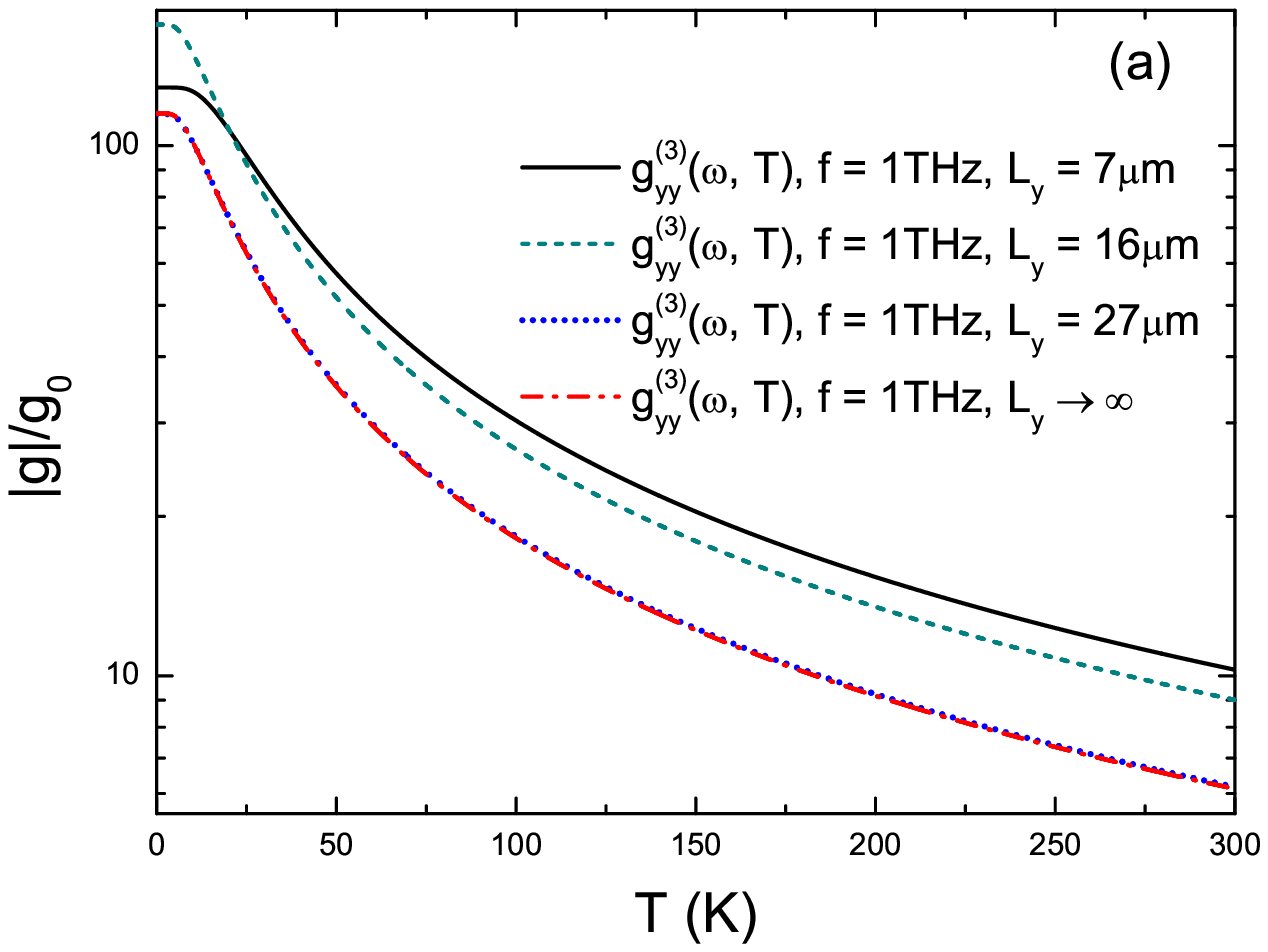}}
\hfil
\subfloat{\label{fig:2b}\includegraphics[width=2.5in]{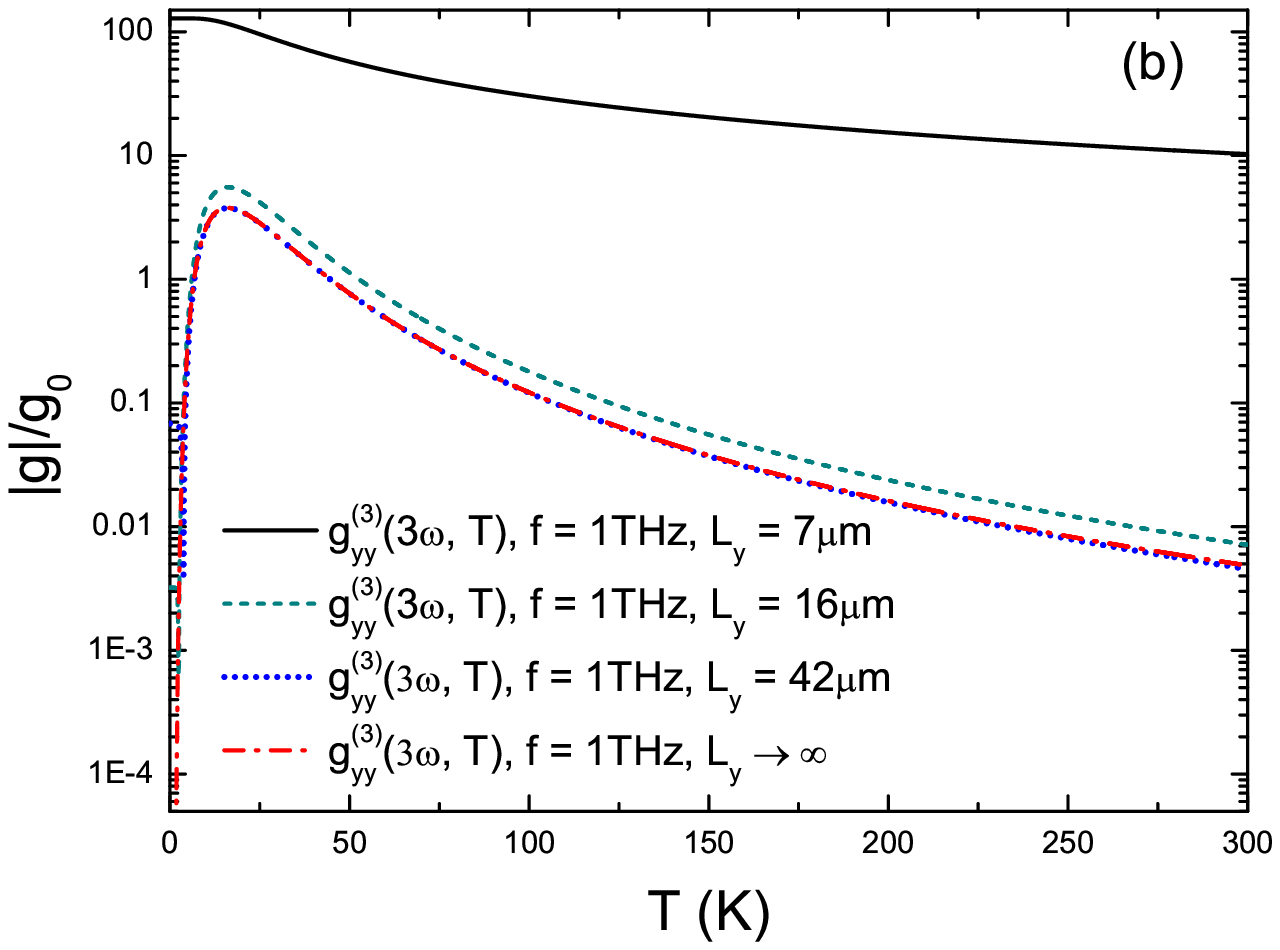}}
\hfil
\subfloat{\label{fig:2c}\includegraphics[width=2.5in]{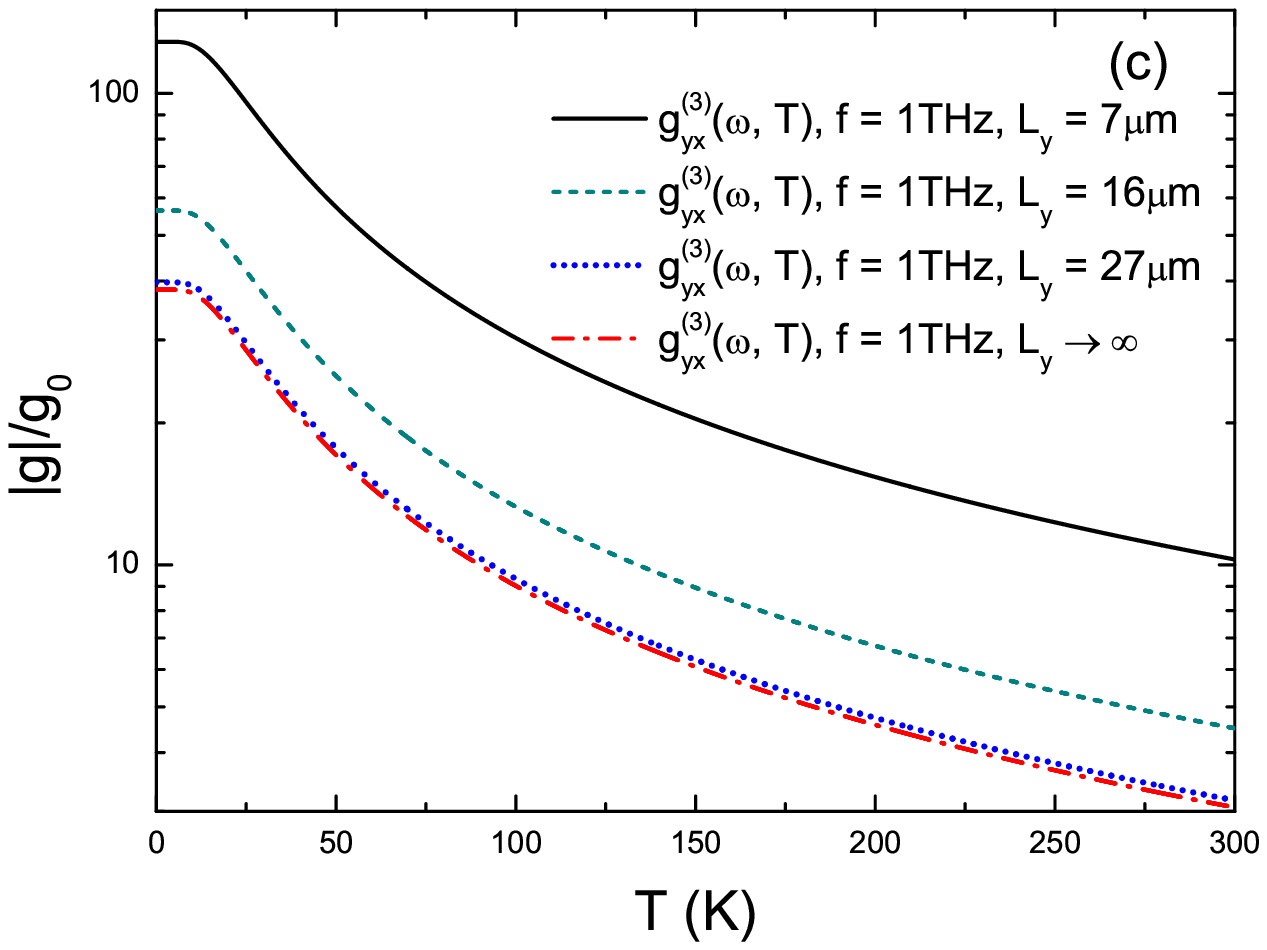}}
\hfil
\subfloat{\label{fig:2d}\includegraphics[width=2.5in]{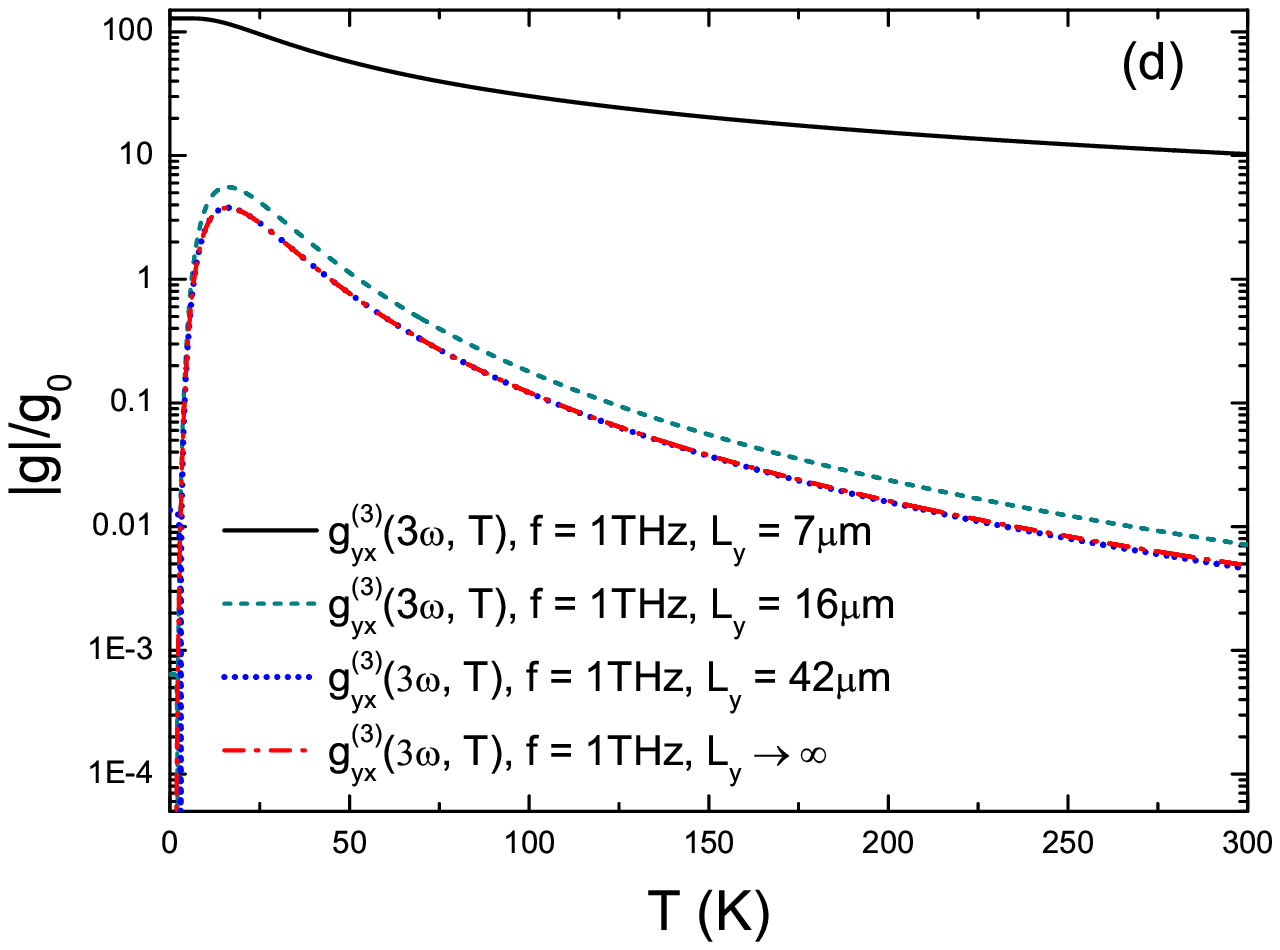}}
\caption{Magnitude of the third-order nonlinear conductances in acGNR20 for
various nanoribbon lengths as a function of temperature $T$. a) isotropic Kerr
conductance, b) isotropic third-harmonic conductance, c) anisotropic Kerr
conductance, and d) anisotropic third-harmonic conductance. For all plots, $f =
\SI{1}{THz}$, $E_F=0$, $L=0$, and $f_{\Gamma}=\SI{0.024}{THz}$
($f_{\Gamma}\to 0$) for nanoribbons of finite (infinite) length.}
\label{fig:2}
\end{figure*}
In Fig. \ref{fig:2} we plot the temperature dependence of the isotropic and
anisotropic AC conductances for several nanoribbon lengths $L_y$ with $L \to 0$. The frequency
of the applied field is $f=\SI{1}{THz}$. For the Kerr conductances as
shown in Figs. \ref{fig:2a} and \ref{fig:2c}: when $L_y=\SI{7}{\micro m}\, (M =
7)$, the conductance is dominated by
the $N(\omega)$ term; when $L_y=\SI{16}{\micro m}\, (M = 16)$, both the
$N(\omega/2)$ and $N(\omega)$ terms contribute; and when $L_y=\SI{27}{\micro
m}$, we arrive at the critical length and the conductance is nearly the same as
that for the infinitely-long nanoribbon for $T\geqslant\SI{0}{K}$. 
For the third-harmonic conductances as
shown in Figs. \ref{fig:2b} and \ref{fig:2d} we observe similar behavior:  when
$L_y=\SI{7}{\micro m}\, (M =
7)$, the conductance is dominated by
the $N(\omega)$ term; when $L_y=\SI{16}{\micro m}\, (M = 16)$, the
$N(\omega/2)$, $N(\omega)$, and $N(3 \omega/2)$  terms contribute; and when
$L_y=\SI{42}{\micro m}$, we arrive at the critical length and the conductance is
nearly the same as that for the infinitely-long nanoribbon for $T\geqslant\SI{5}{K}$. \par
\begin{figure*}[!t]
\centering
\subfloat{\label{fig:3a}\includegraphics[width=2.5in]{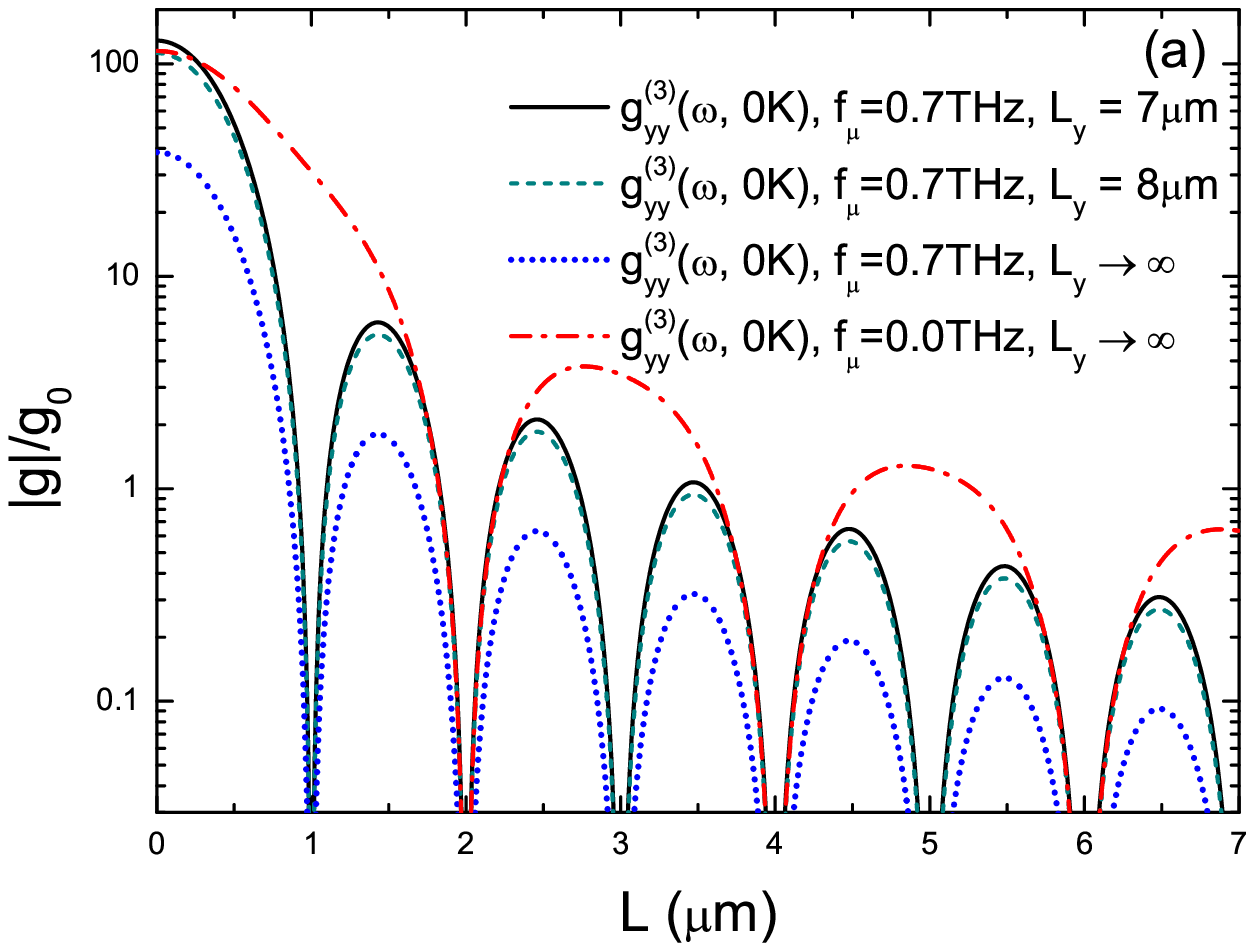}}
\hfil
\subfloat{\label{fig:3b}\includegraphics[width=2.5in]{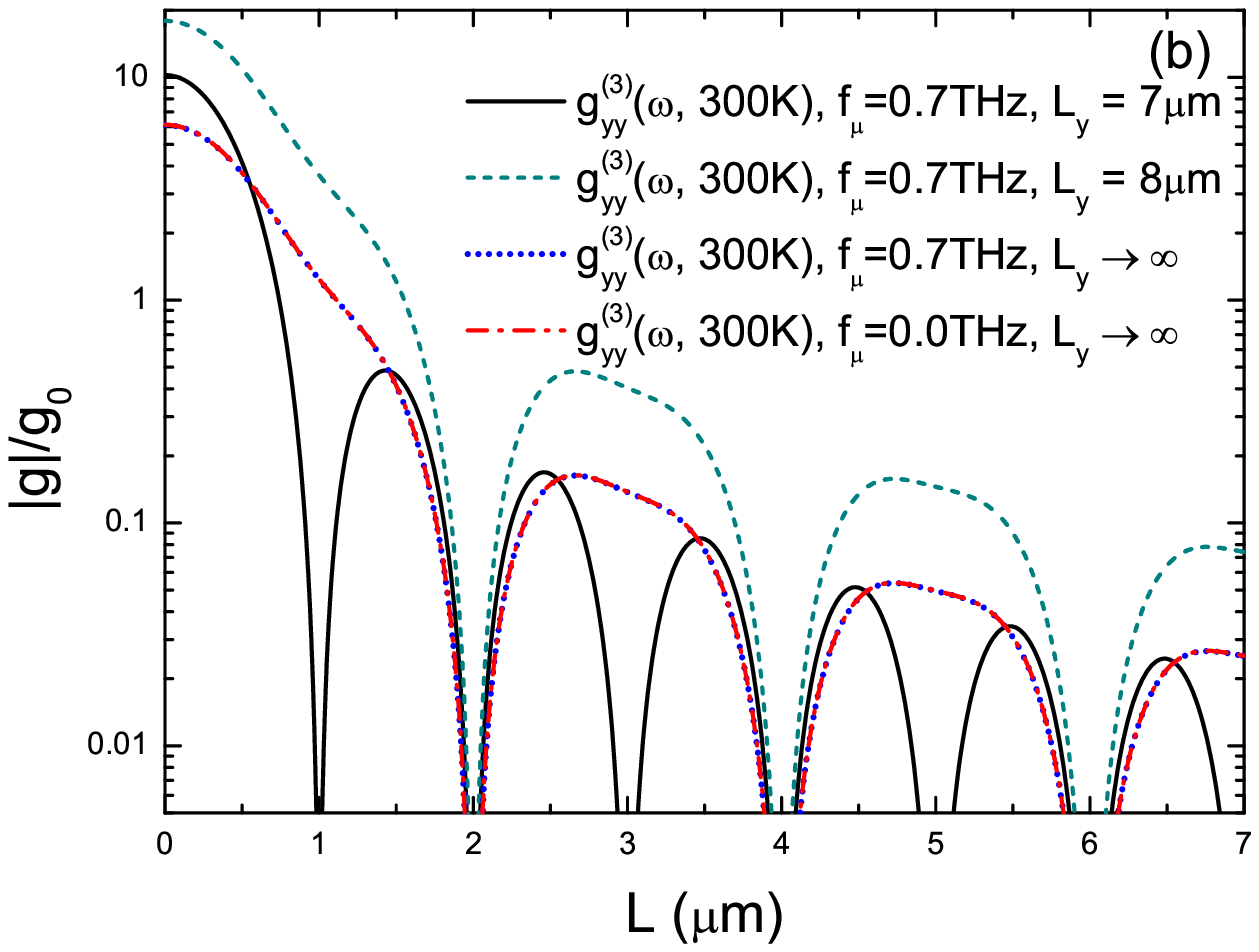}}
\hfil
\subfloat{\label{fig:3c}\includegraphics[width=2.5in]{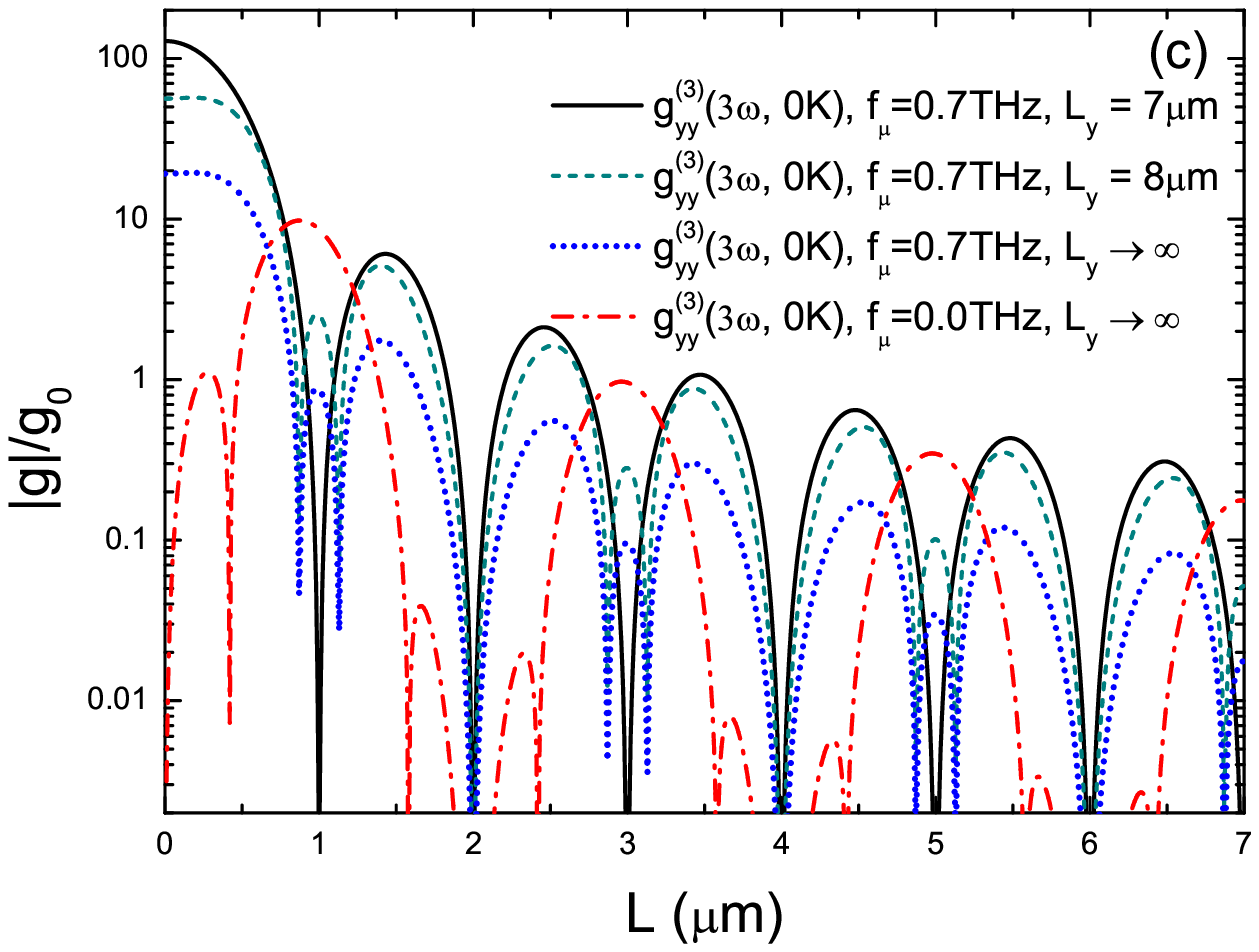}}
\hfil
\subfloat{\label{fig:3d}\includegraphics[width=2.5in]{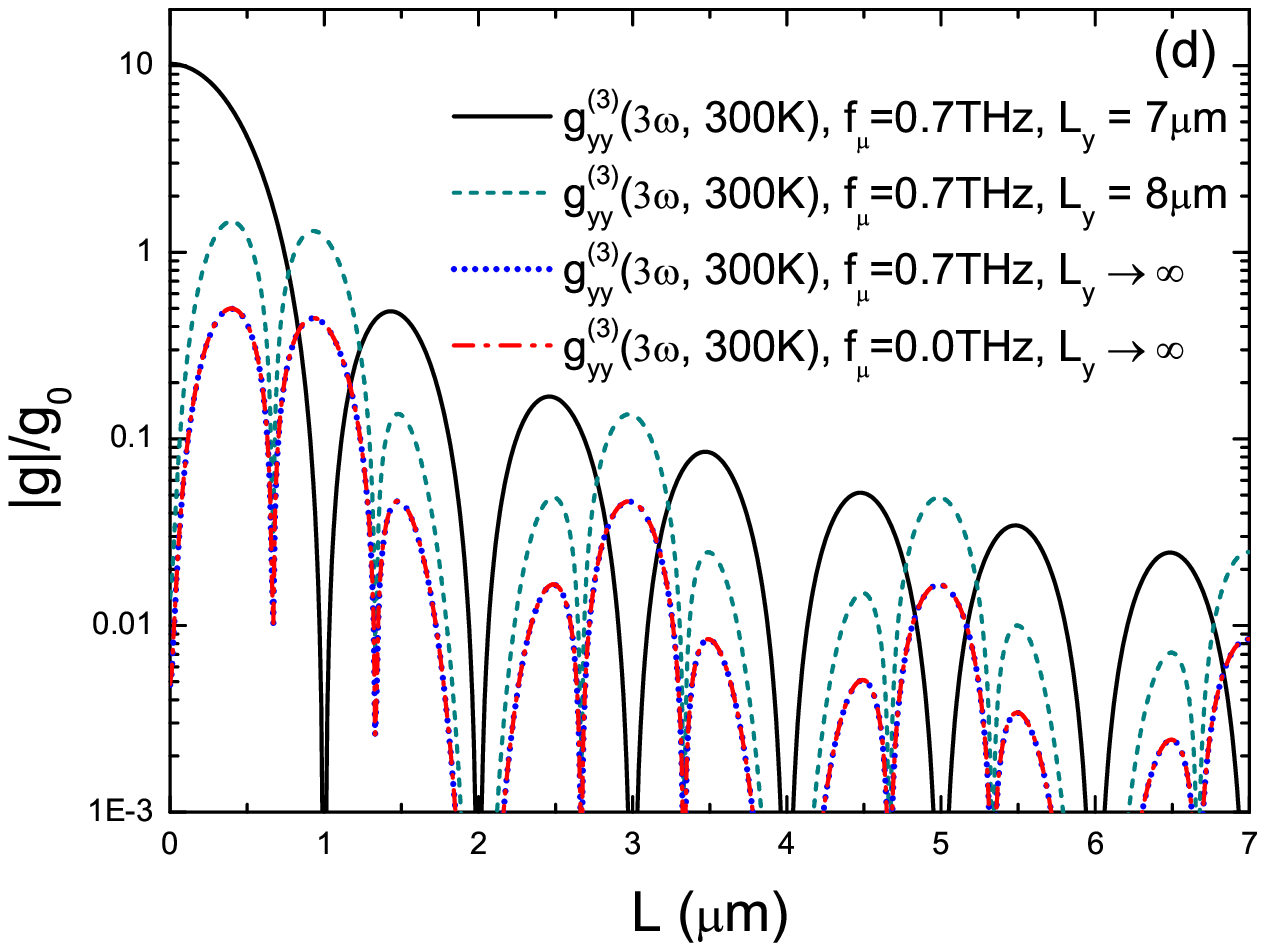}}
\caption{Magnitude of the third-order nonlinear conductances in acGNR20 for
various nanoribbon lengths and Fermi levels as a function of illumination length $L$. a)
isotropic Kerr conductance, b) isotropic third-harmonic conductance, c)
anisotropic Kerr conductance, and d) anisotropic third-harmonic conductance.
For all plots, $f=\SI{1}{THz}$, $f_{\mu}=E_F/h$ and $f_{\Gamma}=\SI{0.024}{THz}$
($f_{\Gamma}\to 0$) for nanoribbons of finite (infinite) length.}
\label{fig:3}
\end{figure*}
It has been shown that the local AC current response depends on the spatial
profile of the applied electric field for quasi 1D quantum wires
\cite{mavsek89,velicky89ac,kramer90coherent}.
In Fig. \ref{fig:3}, we plot the isotropic nonlinear
conductances as a function of the illumination length $L$.
For the Kerr conductance plotted in Figs. \ref{fig:3a} and \ref{fig:3b} we note a series
of antiresonances in the magnitude of the conductance. For the intrinsic 
nanoribbon, these antiresonances
occur when the zeros of the illumination factor $S(\omega, L)$ are located
at the same frequency as the states coupled by the excitation field of frequency
$f$ (those having non-negligible contributions from $N(\omega_y)$).
For the Kerr conductance of the intrinsic nanoribbon with $L_y \to \infty$,
both the $N(\omega/2)$
and $N(\omega)$ terms contribute to the conductance, resulting in an
antiresonance spacing determined by setting the zeros in the illumination factor
equal to $\omega/2$. This results in the set of antiresonances at $L = \SI{2}{\micro m}, 
\SI{4}{\micro m}, ...$.
For the extrinsic case, there are two mechanisms contributing to the
antiresonances: \textit{1)} the zeros in the illumination factor, and
\textit{2)} state blocking due to the Fermi level offset. For example, in Fig.
\ref{fig:3a} ($T = \SI{0}{K}$),
for the extrinsic nanoribbons transitions at $\omega/2$ are
completely blocked,
and therefore only the $N(\omega)$ term contributes, resulting in an antiresonance spacing
determined by setting the zeros in the illumination factor equal to $\omega$.
This results in the set of antiresonances at $L = \SI{1}{\micro m}, \SI{2}{\micro m}, ...$
for both $M$ even and odd. In contrast, for Fig. \ref{fig:3b} ($T = \SI{300}{K}$), for the $M$ odd case, the $N(\omega/2)$ contribution is
negligible and we obtain a similar result as for the $T =\SI{0}{K}$ case.
However, for $M$ even, the $N(\omega/2)$ term is not completely blocked, and
as a result (setting the zeros in the illumination factor equal to $\omega/2$),
we obtain antiresonances at $L = \SI{2}{\micro m}, \SI{4}{\micro m}, ...$.

For the third-harmonic conductance plotted in Figs. \ref{fig:3c} and
\ref{fig:3d}, the behavior is even richer than for the Kerr conductance. While
the set of primary antiresonances follows the discussion above, a pair of sidelobe
resonances surrounding each primary resonance is also observed. These sidelobe
resonances result from the fact that the $N(\omega/2)$ and $N(3 \omega/2)$ terms
in the expression for the conductance (\ref{eq:gACy}c) have the opposite sign of the $N(\omega)$
term. Thus, for certain non-zero values of the illumination factor $S(\omega,
L)$, the positive and negative thermal factor contributions exactly cancel and
the sidelobe antiresonances manifest themselves. Because the exact value of the
thermal factor functions change with temperature, the locations of these sidelobe
antiresonances also shift with temperature. This effect can also be observed in
Figs. \ref{fig:1b} and \ref{fig:1d}.

We also note here that for $T = \SI{0}{K}$, and $L_y \to \infty$, with a
uniform illumination factor $S(\omega, L) = 1$ ($L \to 0$) the
third-harmonic conductance is always zero for intrinsic nanoribbons and only
non-zero over a limited frequency range for extrinsic nanoribbons
($2|E_F|/3\leqslant \hbar \omega \leqslant2|E_F|$) \cite{arxiv1}. By extending
the illumination range $L$ (and therefore modulating the illumination factor)
it is possible to enhance the third-harmonic conductance in these conditions so
that for particular illumination lengths $L$, the third-harmonic conductance
becomes of the order of the Kerr conductance magnitude.

In the interest of brevity, we do not plot the anisotropic nonlinear
conductances, but simply note that the expression for the Kerr conductance
contains only an $N(\omega)$
term. As a result, the characteristics of the anisotropic Kerr conductance follows
closely that of the isotropic Kerr conductance with resonances at $L = \SI{1}{\micro m}, \SI{2}{\micro m}, ...$ as discussed above.
On the other hand, the anisotropic third-harmonic conductance contains all of
the richness of its isotropic counterpart. \par
\begin{figure*}[!t]
\centering
\subfloat{\label{fig:4a}\includegraphics[width=2.5in]{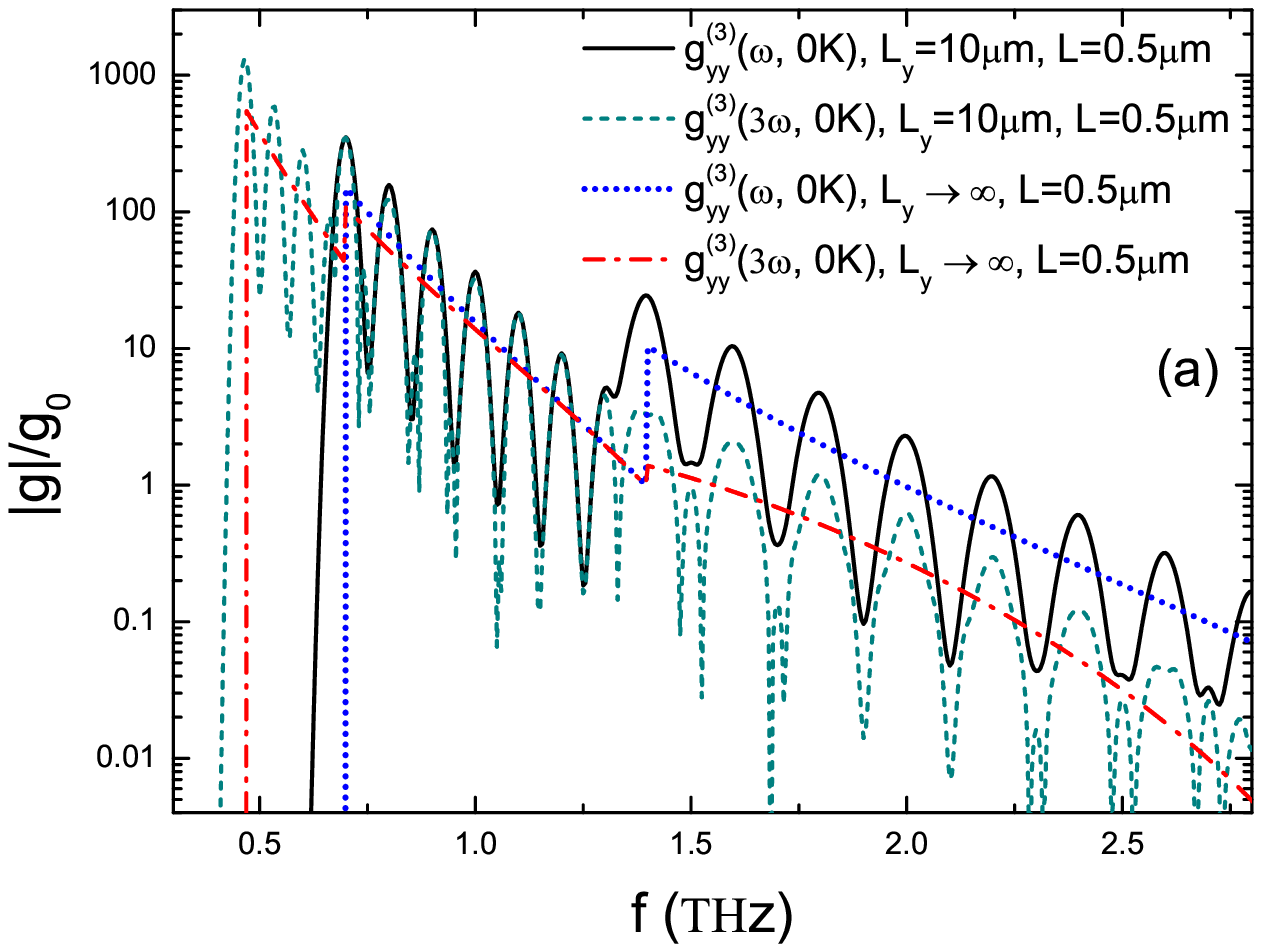}}
\hfil
\subfloat{\label{fig:4b}\includegraphics[width=2.5in]{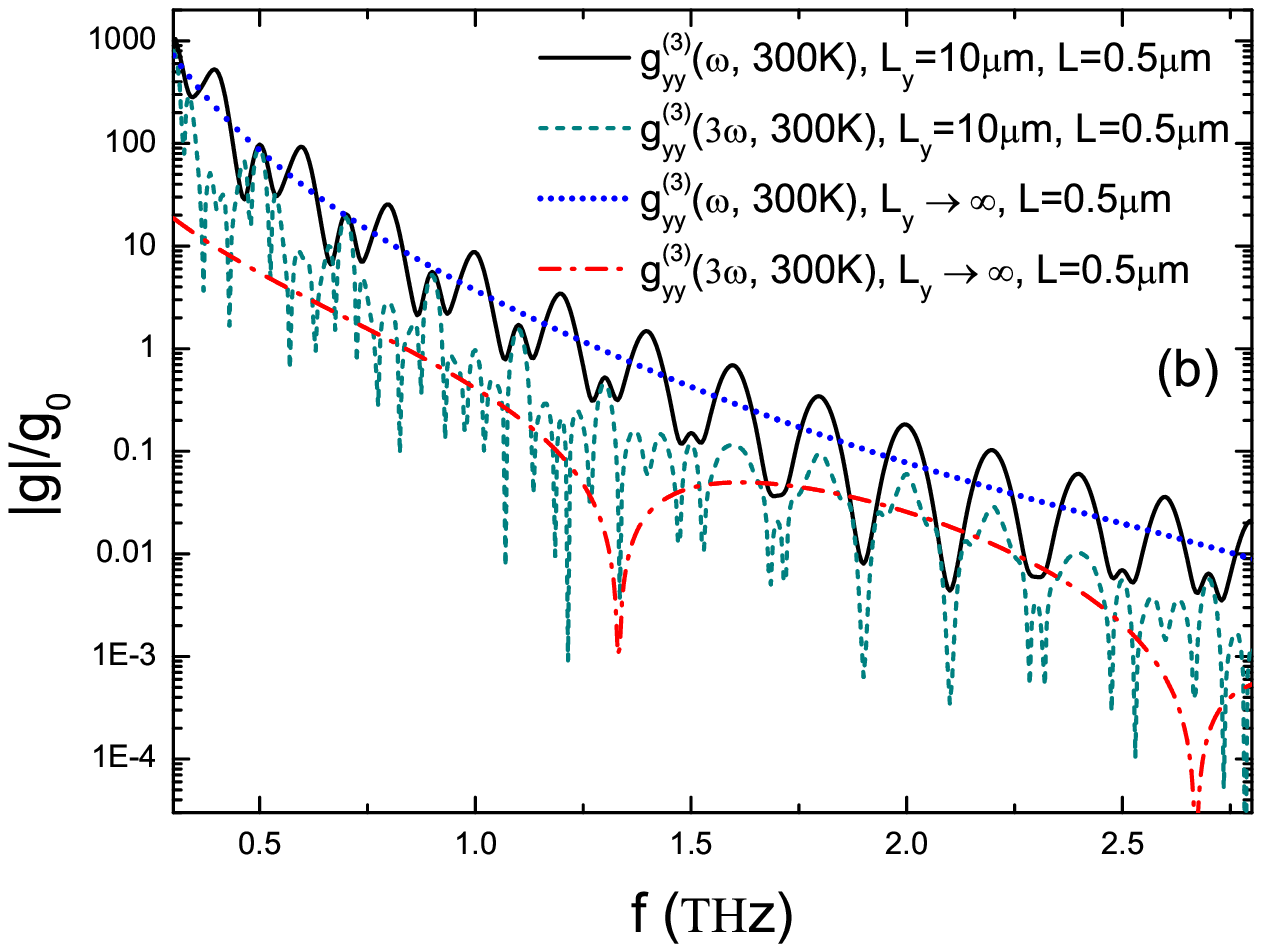}}
\hfil
\subfloat{\label{fig:4c}\includegraphics[width=2.5in]{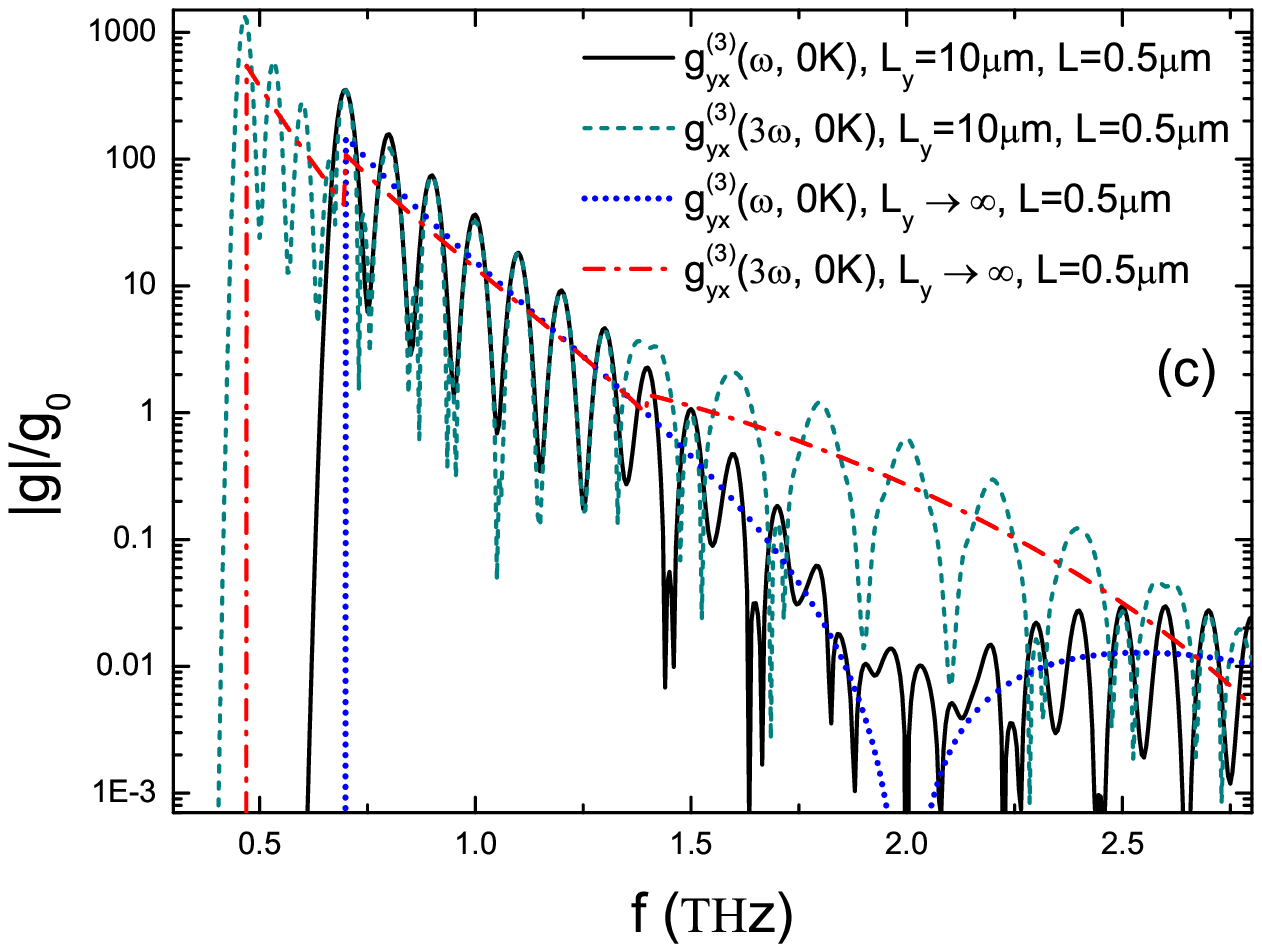}}
\hfil
\subfloat{\label{fig:4d}\includegraphics[width=2.5in]{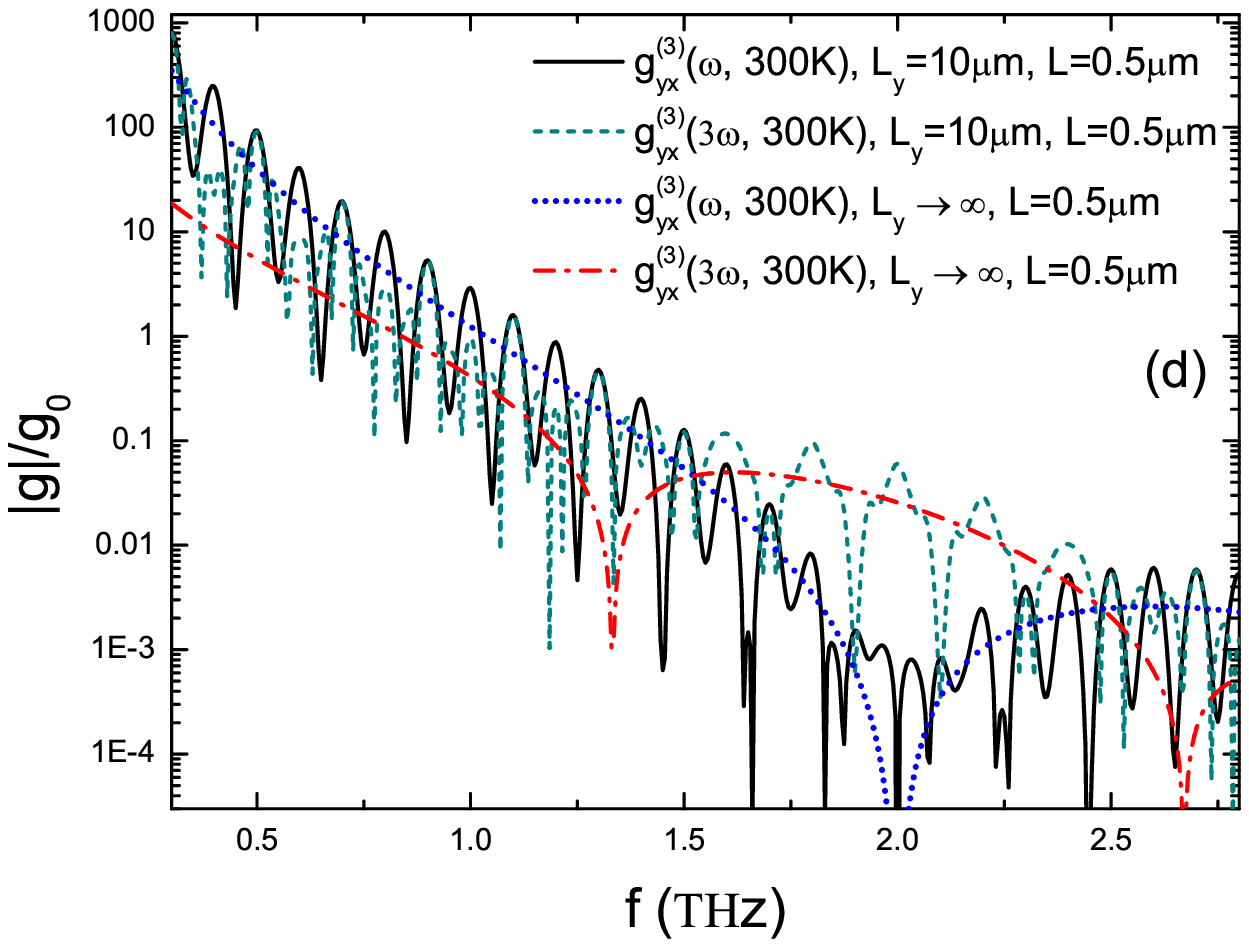}}
\caption{Magnitude of the third-order nonlinear conductances in acGNR20 for
various nanoribbon and illumination lengths, and temperatures as a function of excitation
frequency $f$. a) isotropic Kerr and third-harmonic conductances at $T = \SI{0}{K}$, b) isotropic Kerr and third-harmonic conductances at $T = \SI{300}{K}$, c) anisotropic Kerr and third-harmonic conductances at $T = \SI{0}{K}$, and d) anisotropic Kerr and third-harmonic conductances at $T =
\SI{300}{K}$. For all plots, $E_F/h=\SI{0.7}{THz}$
and $f_{\Gamma}=\SI{0.024}{THz}$
($f_{\Gamma}\to 0$) for nanoribbons of finite (infinite) length.}
\label{fig:4}
\end{figure*}

Fig. \ref{fig:4} illustrates the overall impact of the various quantum size
effects we have discussed above on the magnitude of the third-order Kerr and
third-harmonic conductances. In Figs. \ref{fig:4a} and \ref{fig:4b}, we plot the
isotropic Kerr and third-harmonic conductances for several values of nanoribbon
length $L_y$, temperature, and excitation frequency.
The oscillatory nature
of these curves highlights the interplay between the thermal factor (and thermal
factor cancellation in the case of the third-harmonic conductance), state
blocking, and the illumination factor. While the overall envelope of the $L_y =
\SI{10}{\micro m}, L = \SI{0.5}{\micro m}$
conductances decay with increasing frequency as expected (generally following
the results for the $L_y \to \infty, L = \SI{0.5}{\micro m}$ case), there is clearly a
richness in the detailed oscillatory
behavior governed by the exact characteristics of the illumination and
sample geometries. Similar effects are noted in Figs. \ref{fig:4c} and
\ref{fig:4d} for the anisotropic nonlinear conductances as well.
It is also useful to point out here that it should be possible to modify this
dynamical behavior by apodizing\cite{bornwolf} the applied THz electric field. For example, by using
a Gaussian-apodized profile for the applied electric field, it should be possible to
eliminate the antiresonances induced by the illumination factor $S(\omega, L)$.
This would significantly reduce the oscillatory character of the results
presented in Fig. \ref{fig:4}.

In summary, the results described above place important constraints on the
development of metallic acGNR THz devices. For a fixed excitation field
frequency, the nanoribbon length $L_y$, illumination factor
$S(\omega, L)$, intrinsic broadening $\Gamma_\omega$,
and carrier distribution $(E_F)$
will need to be carefully considered based on a particular device application.
For example, by appropriate choice of these parameters, it is possible to use an acGNR to
generate third-harmonic radiation at $T = \SI{0}{K}$, whereas for
other sets of parameters, the third-harmonic component is zero.
While it is beyond the scope of this paper to delve into such
design details, we note that the results presented here will guide
the designer toward an optimal design. The efficient THz nonlinear response of
acGNR described above provides much promise
toward the development of devices such as polarization switches,
modulators, and efficient background-free third-harmonic generators.

\section{Conclusion}\label{Conclusions}
In this paper, we describe the results of detailed calculations of the quantum size effects on
the nonlinear third-order conductances of acGNR. We report that novel effects due
to both the size and spectral broadening of the nanoribbon, as well as the spatial profile of
the excitation field, are important in determining the nonlinear response of
acGNR. We compute the THz third-order nonlinearities of a thin,
finite-length metallic acGNR. Our calculations show that there is a transition between
quantum dot-like behavior for small $L_y$ and quasi-continuum behaviour as
$L_y$ increases. The boundary between these two regimes is shown to be a
function of the broadening of the Dirac spectrum of the nanoribbon.
Additionally, we observe that the nonlinear response in metallic acGNR is
strongly dependent on the shape of the spatial profile of the THz
excitation field. By carefully choosing the spatial profile,
it is possible to optimize the third-order nonlinearities for a particular
excitation frequency. In the results presented above, we present a detailed analysis of
the features of the nonlinear spectral response due to these mechanisms.

Finally, we note two recent reports of the synthesis of high quality, ultrathin acGNR with
widths $L_x < \SI{10}{nm}$\cite{kimouche2015ultra,jacobberger2015direct}.
The recent advent of this capability to fabricate acGNR underscores the importance
of a complete understanding of the underlying nonlinear physics of these structures.
AC transport in quasi 1D quantum wires is crucial for high speed quantum wire
based integrated circuits \cite{cheng11}.
The current work contributes to this understanding by demonstrating that acGNR have
large nonlinearities that can be optimized through careful choice of design
parameters such as the nanoribbon dimensions and the spatial profile of the excitation
field.

% or


%\appendices


%\section{}

%\section*{Acknowledgment}


\ifCLASSOPTIONcaptionsoff
  \newpage
\fi

\end{document}